\documentclass[%
reprint,
amsmath,amssymb,
aps,
prb,
]{revtex4-2}

\usepackage{graphicx}
\usepackage{dcolumn}
\usepackage{bm}
\usepackage{siunitx}

\begin{document}
	
	\preprint{APS/123-QED}
	
	\title{Gentle tension stabilizes atomically thin metallenes}
	\author{Kameyab Raza Abidi}
	\author{Pekka Koskinen}%
	\email{pekka.j.koskinen@jyu.fi}
	\affiliation{%
		NanoScience Center, Department of Physics, University of Jyväskylä, 40014 Jyväskylä, Finland
	}%
	
	\date{\today}
	
	\begin{abstract}
	Metallenes are atomically thin two-dimensional (2D) materials lacking a layered structure in the bulk form. They can be stabilized by nanoscale constrictions like pores in 2D covalent templates, but the isotropic metallic bonding makes stabilization difficult. A few metallenes have been stabilized but comparison with theory predictions has not always been clear. Here, we use density-functional theory calculations to explore the energetics and dynamic stabilities of $45$ metallenes at six lattices (honeycomb, square, hexagonal, and their buckled counterparts) and varying atomic densities. We found that of the $270$ different crystalline lattices, $128$ were dynamically stable at sporadic densities, mostly under tensile strain. At the energy minima, lattices were often dynamically unstable against amorphization and the breaking down of metallene planarity. Consequently, the results imply that crystalline metallenes should be seen through a novel paradigm: they should be considered not as membranes with fixed structures and lattice constants but as yielding membranes that can be stabilized better under tensile strain and low atomic density. Following this paradigm, we rank the most promising metallenes for 2D stability and hope that the paradigm will help develop new strategies to synthesize larger and more stable metallene samples for plasmonic, optical, and catalytic applications.
	\end{abstract}

\maketitle

The early days of two-dimensional (2D) materials research were greatly influenced by the easiness of graphene exfoliation \cite{graphene}. Since then, synthesis methods have significantly evolved, producing 2D materials beyond graphene with unique applications \cite{beyondgraph, Graphenebeyond, TMD, application2D}. This evolution has been enabled by 2D materials' strong covalent in-plane bonding and weak van der Waals out-of-plane bonding. 

In this respect, metallenes are fundamentally different. Their metallic, non-directional bonding favors 3D aggregation, makes exfoliation challenging, and renders large, free-standing membranes unstable \cite{review4, review3}. Small metallene patches have nevertheless been successfully stabilized by various means. One example is the stabilization in the pores of 2D covalent templates as side products from migrating metal residues on surfaces \cite{zhao_free-standing_2014, 2DAu, 2Dtin}. The second example is to use electron irradiation to ionize Se atoms from MoSe$_2$ transition metal dichalcogenides to make patches of 2D Mo \cite{2DMo} or Ag atoms from AuAg alloy to produce a small suspended Au monolayer \cite{2DGold}. A third example is the etching of Ti$_3$C$_2$ of nanolaminated Ti$_3$AuC$_2$ to create large-area sheets of 2D Au, goldene \cite{Goldene2024}. We emphasize that, like in the experiments, our focus here is on \emph{suspended} crystalline metallenes, not supported or amorphous ones. The stabilization is challenging, but metallic bonding is the very feature that makes these materials unique among other 2D materials \cite{gold_in_graph, plentyofmotion, Liquid_coexist} and attractive for plasmonic, electronic, quantum dot, energy, biomedical, and catalytic applications \cite{metallene_2, metallene_1, metallene_3}. This attractiveness has motivated a fair amount of research \cite{review1, review2, review3, review4, review5}.

\begin{figure}[t!]
\centering
\includegraphics[width=0.8\columnwidth]{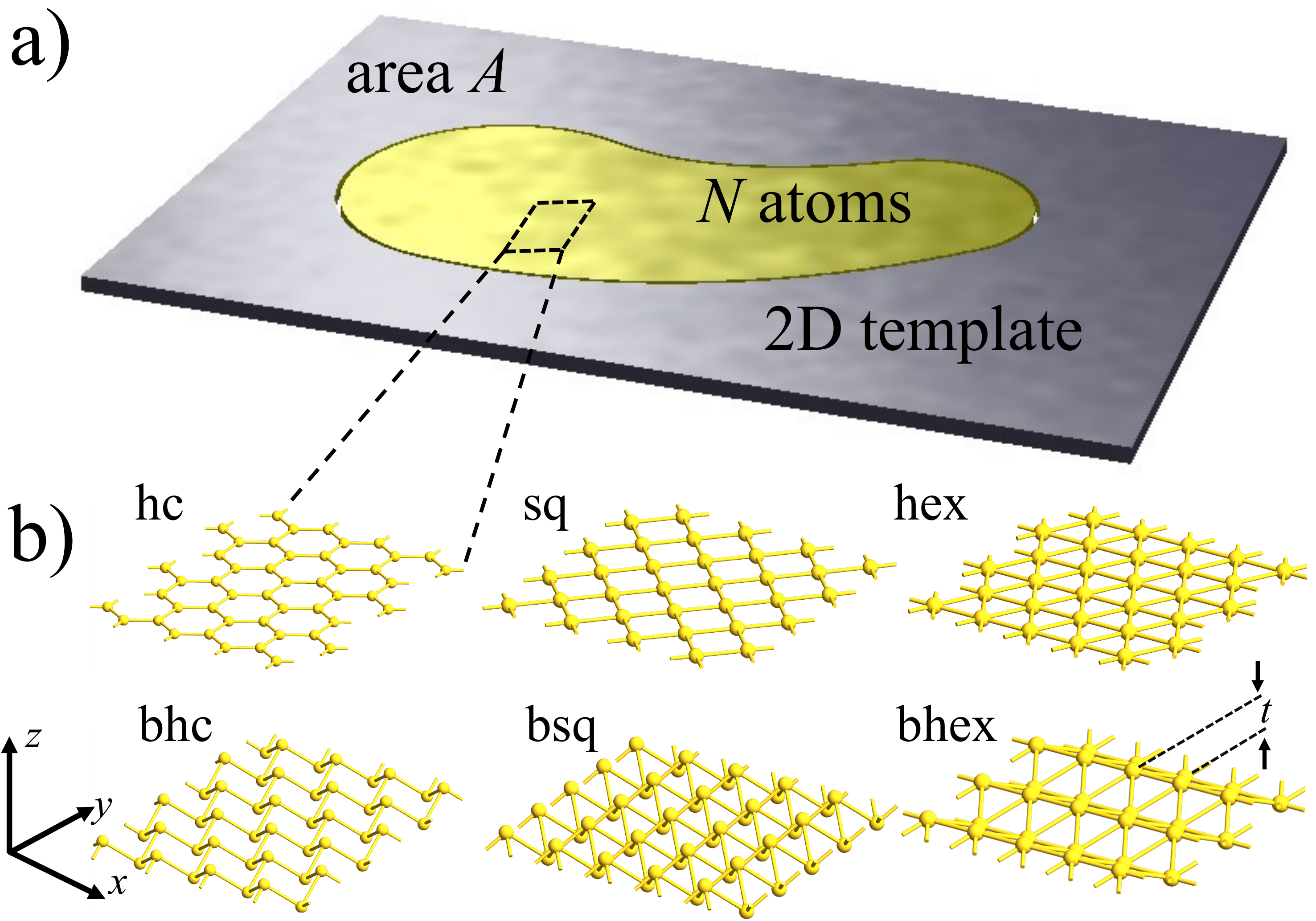}
\caption{Illustrating the proposed stability paradigm of metallenes. a) An $N$-atom, suspended metallene patch is stabilized by the interface with the pore of area $A$ in a 2D template such as graphene. The patch can be under stress, whereby the atom density $N/A$ may or may not correspond to the energy minimum of the isolated metal monolayer. b) The honeycomb (hc), square (sq), and hexagonal (hex) lattice monolayers investigated in this Communication, together with their buckled counterparts (bhc, bsq, and bhex). The thickness of the buckled monolayer is $t$. For the purposes of investigating general stability trends, our approach is based on \emph{focusing on the lattice properties with a local viewpoint} and \emph{not} considering finite patches with explicit interfaces.}
\label{fig:systems}
\end{figure}

Meanwhile, density-functional theory (DFT) simulations have played a significant role in predicting diverse properties of metallenes \cite{nevalaita_atlas_2018, ren_magnetism_2021, pawar_thickness, nevalaita_free-standing_2020, nevalaita_stability_2019, ono_comprehensive_2021}. The dynamical stability analysis of monolayers has been a particularly valuable procedure guiding the experiments \cite{ono_dynamical_2020}. So far, simulations have aimed to align with the traditional paradigm of covalent 2D materials: they have been considered periodic 2D bulk with optimized lattice constant corresponding to the energy minimum. Unfortunately, comparing simulations with a limited number of metallene samples has been ambiguous regarding both lattice symmetry and bond lengths \cite{2dironfree, Fe_Carbide_1, Fe_Carbide_2, 2DCr, 2Dtin, Goldene2024}. This ambiguity is understandable, as the number of metal atoms $N$ in the pore may vary, while the pore area $A$ remains fixed (Fig.~\ref{fig:systems}a). Thus, given their high ductility with respect to the covalent template \cite{plentyofmotion}, metal patches \emph{could be strained, making stability not intrinsic to the metal} but dependent on the atomic density $N/A$.

Therefore, in this Communication, we address the question: \emph{to what extent do the dynamical stabilities of various metallene lattices depend on atomic densities?} We address this question by DFT calculations of the energetic and dynamic stabilities of six metallene lattices (honeycomb, square, hexagonal, and their buckled counterparts) for $45$ elements as a function of atomic density, quantified in terms of specific area $a=A/N$. Unlike covalent 2D materials, we found that metallene stability is highly sensitive to atomic density. Out of the $270$ monolayers, $128$ were stable at sporadic densities, dominantly under tensile stress, and only occasionally at zero stress. Supported by molecular dynamics simulations, we also found that different lattices can coexist at a density where any single lattice would be unstable. Juxtaposed with metallenes' softness and ductility\cite{plentyofmotion}, these results suggest a new stability paradigm that abandons fixed lattice constants and considers metallenes as yielding patches that can be stabilized better under tensile strain. In other words, the energy of the patch is not minimized separately but together with the total energy of the stabilizing interface.

To address the above question, we investigated the energetics and dynamical stabilities of the systems (Fig.~\ref{fig:systems}) by density-functional simulations \cite{kohn_self-consistent_1965, QuantumATK}. We used the PBE exchange-correlation functional and an LCAO basis set \cite{PBE}, a level of theory that has proven robust for metallenes and that suffices well for our purposes of investigating general trends in stabilities and geometries \cite{OptimalDFT}. The lattices were modeled by minimal cells periodic in the $xy$-plane using $13\times 13\times 1$ $k$-point sampling (Fig.~\ref{fig:systems}b). The buckled lattices had initial-guess thicknesses of $t = d\sqrt{|\varepsilon|\,(-|\varepsilon|+2)}$, where $d$ is the equilibrium bond length of the corresponding flat lattice and $\varepsilon$ is the biaxial strain. For details of the computational methods, see Electronic Supplementary Information (ESI).  

The $N$ atoms in the periodic cell of area $A$ were then optimized for each element and lattice type, yielding the cohesive energy per atom $E_\text{coh}^L(a)=E^L(a)/N-E_\text{free}$ as a function of specific area $a=A/N$, where $E^L(a)$ is the optimized energy of lattice $L$ and $E_\text{free}$ is the energy of a single atom inside a $15$-\AA\ cube. The dynamical stabilities were analyzed by calculating phonon dispersion spectra and examining the absence of phonon modes with imaginary frequencies \cite{phonon}. This systematic approach resulted in a fair number of calculations: a total of $45$ elements with $6$ lattices were optimized for $40-70$ different values of $a$, with complete phonon dispersion spectra calculations for most configurations, giving some $\gtrsim 15000\;$ phonon calculations in total.

\begin{figure}[t!]
\centering
\includegraphics[width=0.8\columnwidth]{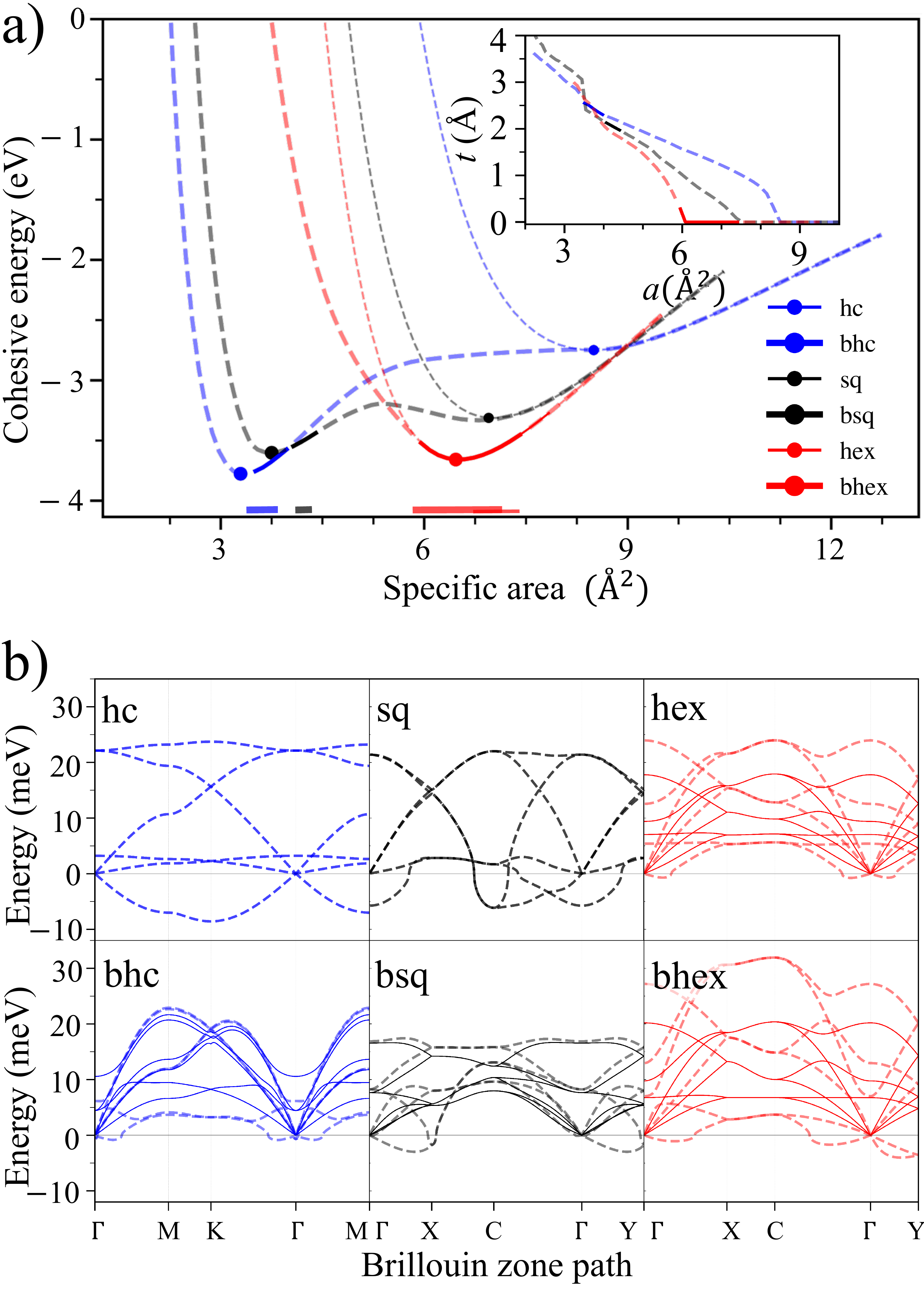}

\caption{Dependence of 2D Au dynamical stability on atom density. a) The cohesive energies per atom of the six lattices (dashed lines) and the dynamically stable intervals (solid segments; also highlighted above abscissa). The buckled lattices become equivalent to the flat ones at the onset of buckling. The energy scale is from gas phase atoms (zero) to 3D bulk cohesion (minimum). Dots denote the energy minima. Inset: the thickness of buckled lattices as a function of atom density. b) Phonon dispersion spectra of Au with different lattices and densities. Dynamically stable configurations (solid lines; middle of stable segments) have phonon modes with positive energies, whereas dynamically unstable ones (dashed lines; at energy minima or the nearest unstable configuration at a smaller density) also have phonon modes with negative energies, implying imaginary frequencies and saddle points in the potential energy surface.}
\label{fig:Au} 
\end{figure}

To discuss the main features of the results, let us use Au as an illustrative example. All six lattices show typical cohesion energy curves as a function of $a$ (Fig.~\ref{fig:Au}a). Buckled honeycomb and hexagonal lattices bind strongest, while honeycomb and square bind weakest. All lattices have one energy minimum, except for buckled square, which has two. The cohesive energies are large, $88$~\%\ (hexagonal) and $91$~\%\ (buckled honeycomb) of the 3D bulk cohesion.

However, only a fraction of the configurations are dynamically stable (solid line segments in Fig.~\ref{fig:Au}a). Four of the six lattices are stable at some densities (bhc, bsq, bhex, and hex), and two are never stable (hc and sq). Overall, dynamically stable 2D Au can be found at sporadic area intervals of $3.5\textup{--}4.0$~\AA$^{2}$ (bhc), $4.1\textup{--}4.4$~\AA$^{2}$ (bsq), and $6.7\textup{--}7.8$~\AA$^{2}$ (hex). Such restricted stability \emph{starkly contrasts covalent 2D materials} such as graphene, silicene, or hexagonal boron nitride (hBN) that are stable over considerably larger density ranges (Fig.~S2). 

Phonon dispersion spectra provide the cohesion energy curves with a complementary viewpoint. Near the onset of buckling ($a\approx 6.0$~\AA$^{2}$), hexagonal and buckled hexagonal lattices are near degenerate, but their phonon mode behaviors differ strongly. The buckled hexagonal lattice acquires prominent negative phonon energies, implying acute instability (Fig.~\ref{fig:Au}b). The flat hexagonal lattice, in turn, becomes unstable through modestly negative energies in long-wavelength acoustic modes around the high symmetry points of the Brillouin zone. Most stable lattices become unstable through the gradual emergence of long-wavelength negative-energy phonon modes. We also investigated unstable phonon modes at or near the energy minima of different lattices. In short, it appears that following the unstable phonon modes leads to amorphous structures and the disruption of planarity, a situation not interesting for experiments.

\begin{figure}[t!]
\centering
\includegraphics[width=1.0\columnwidth]{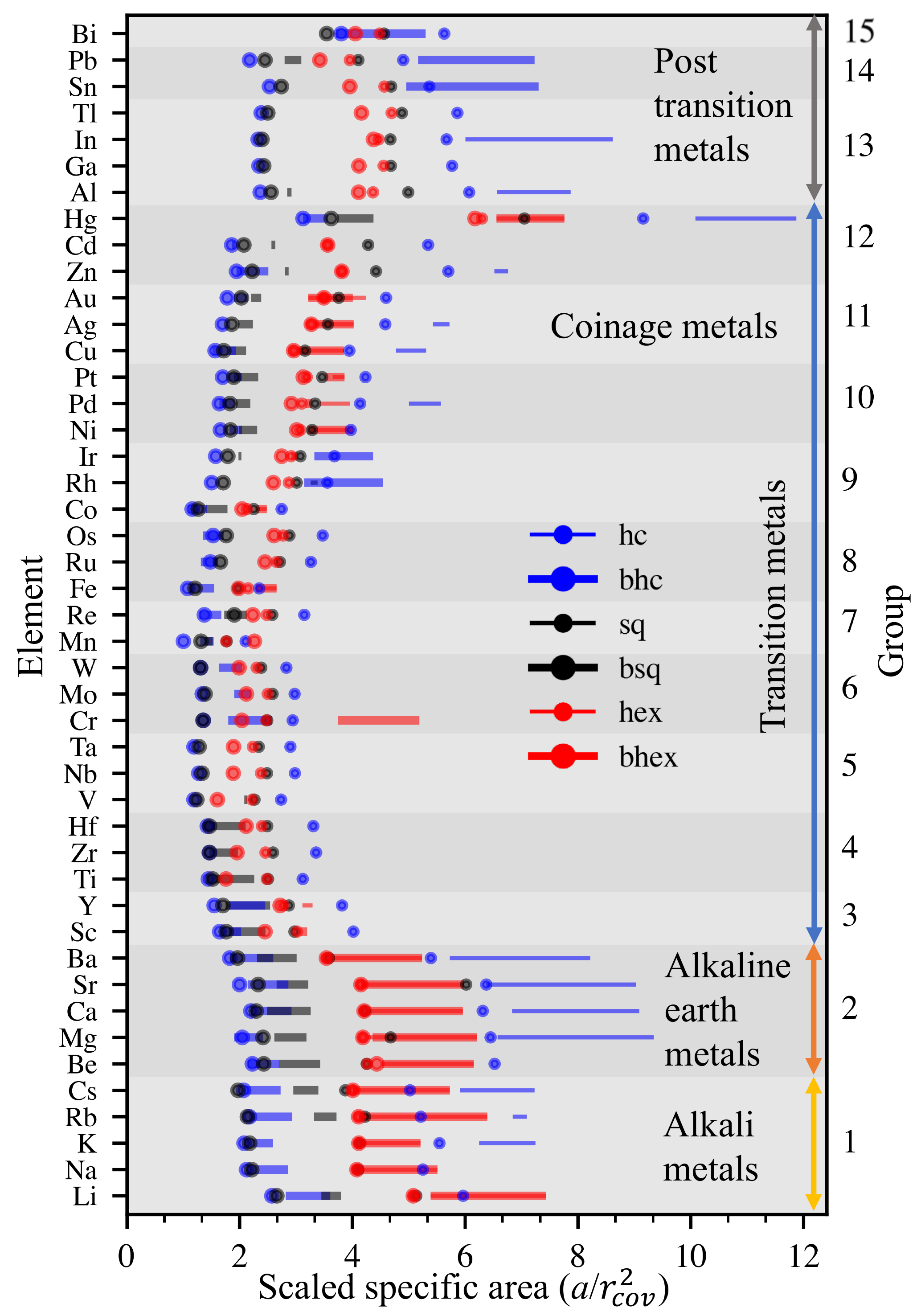}
\caption{Stability trends of $45$ elemental metallenes. Dynamically stable intervals (solid line segments) as a function of the specific areas scaled by the squares of elements' covalent radii. The dots show the cohesive energy minima of each lattice, whether dynamically stable or not.}
\label{fig:all_elements}
\end{figure}

Features similar to Au can also be found in the other $44$ elements. We compare the stabilities of various elements using specific areas scaled by the squares of elements' covalent radii, $a'=a/r_\text{cov}^2$ \cite{Covalent_radii}. This scaling establishes visually organized trends in the positions of the cohesion energy minima of the six lattices and provides an illustrative viewpoint for inspecting the stabilities further (Fig.~\ref{fig:all_elements}).

The best dynamical stabilities occur for alkali and earth alkali metals of periodic table groups $1$ and $2$, Mg showing the most extensive range of stable densities among all elements. Among transition metals (groups 3-12), the most extensive stability ranges occur for early and late transition metals, including coinage metals. In the middle of the transition series, dynamical stability is severely limited. In post-transition metals, the dynamical stabilities vary wildly. Metals are stable at very different densities, and stable intervals are separated by more significant gaps than in groups 1 and 2. Ga and Tl are stable only at singular densities, whereas Hg is stable here and there over an extensive density range. 

In terms of intrinsic material properties, stability is emphasized for simple metals and metals with filled (or nearly filled) d-shells, whose properties are determined by s- and p-electrons. In other words, the directional bonding of d-orbitals decreases the overall 2D stability. Stability correlates negatively with the difference in 3D and 2D cohesion energies, small differences resulting in more stable metallenes \cite{nevalaita_stability_2019}. Moreover, the melting point of 3D bulk, which is caused by directional bonding of partly filled d-shell, correlates negatively with metallene stability.

The results compare well with the experiments done so far. The hexagonal lattice of Au is stable between $6.7\textup{--}7.8$~\AA$^{2}$ ($2.8\textup{--}3.0$~\AA\ bond lengths), which agrees with previous experiments \cite{2DGold,Goldene2024}. The bhc lattices in Sn, Pb, Bi, Cr, Mo, W, Rh, and Ir have two minima, one at higher and one at lower density. Certain experimentally observed bhc lattices appear to favor lower density \cite{Plumbene, Stanene, Bismuthene}. Dynamically stable Fe is found only in bhc, hex, and bhex lattices, supporting the viewpoint that the Fe monolayer lattice reported by Zhao \emph{et al.} \cite{2dironfree} is in carbide and not in elemental form \cite{Fe_Carbide_1, Fe_Carbide_2}. Cr presents a notable exception in the stability among transition metals. It also has been observed experimentally, although the agreement with theory was ambiguous, presumably due to finite size effects \cite{2DCr}.

In particular, the results provide valuable insights to expedite future experiments. Lattices' energy minima are dynamically stable merely occasionally, mostly for hexagonal lattices. As a rule of thumb for all lattices, \emph{metallenes are most likely dynamically stable under gentle tensile strain}, like soap membranes stabilized by a hoop. They are hardly ever dynamically stable under compressive strain. For best stability, metallenes should be seen through a novel paradigm, considering them not as membranes with fixed lattice constants but as yielding membranes that can be stabilized better under gentle tension. 

The results also teach a lesson about metallic bonding in 2D. Honeycomb lattices with covalent bonds can be stretched substantially. The atomic density range [specific area (max-min)/min] for dynamic stability is $43.1$ $\%$ for graphene, $45.5$ $\%$ for silicene, and $39.2$ $\%$ for hBN (Fig.~S2). Such large stability ranges are due to bond directionality supporting lattice symmetry (Fig.~S2). However, metallic bonding is non-directional and nonlocalized in character. For many 2D metals, this character suppresses dynamic stabilities, rendering most of the studied 2D metal lattices to have stable atomic density ranges considerably smaller than covalent 2D materials. For others, lattices break down gradually upon sufficient tensile strain---the non-directionality of metallic bonding cannot preserve lattice symmetry. In other words, the mechanical ductility of metals inevitably comes along with sensitivity for tensile strain: the strain must be gentle.

\begin{figure}[t!]
\centering
\includegraphics[width=0.8\columnwidth]{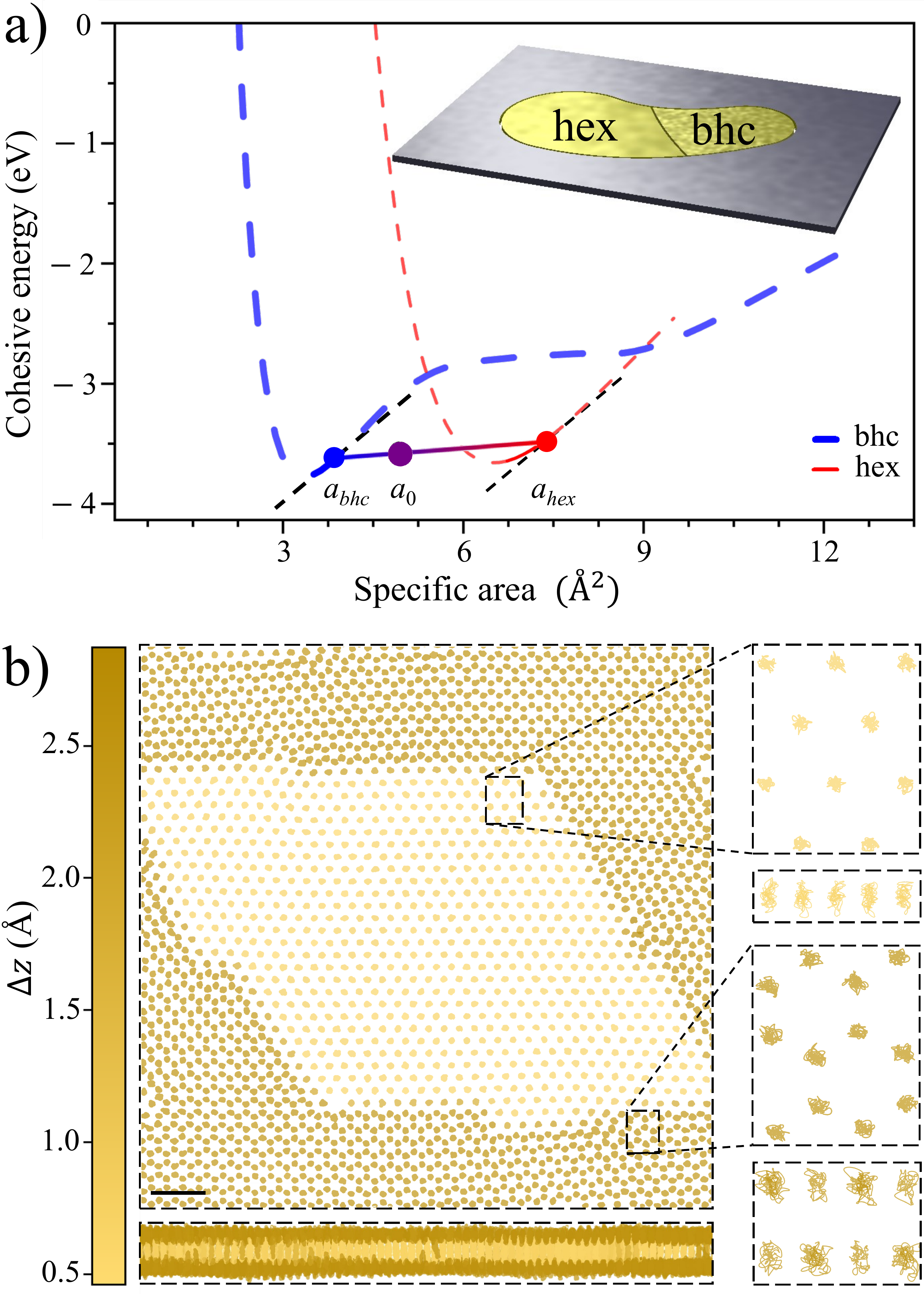}

\caption{Illustrating lattice coexistence in Au. a) Cohesion energy curves of Au for hexagonal and buckled honeycomb lattices (adopted from Fig.~\ref{fig:Au}a). Buckled honeycomb lattice at $a_{bhc}=3.85$~\AA$^{2}$ and hexagonal at $a_{hex}=7.41$~\AA$^{2}$ have the same stress and may exhibit coexistence with $E^\text{coh}_{bhc/hex}(a)$ along the line shown. Inset: schematic of a pore of area $A$ filled by coexisting bhc and hex lattices at $a_0=5.41$~\AA$^2$ mean specific area, with $x=56$~\%\ of atoms in bhc lattice taking a fraction of $xa_{bhc}/a_0=40$~\%\ of the total area $A$, corresponding to the situation in the main figure. b) Coexisting bhc (dark) and hex (light) lattices in a $97.9$~nm$^2$ supercell with $2018$ atoms. The color comes from the vertical thickness $\Delta z$ of the atom's immediate neighborhood. Top and side views of $10$-ps trajectories for atoms within the boxed areas, demonstrating lattice coexistence and dynamical stability, adopted from a $0.25$-ns Langevin molecular dynamics simulation at $300$-K, using a force field with cohesion energy curves similar to those in panel a (see ESI for details).} 
\label{fig:coexistence}
\end{figure}

So far, the focus has been on the stability of a single lattice, but certain lattices may coexist at nearly arbitrary mean density. This phenomenon is familiar from the coexistence of regular 3D phases. Let a pore of area $A$ be filled by $N=N_1+N_2$ atoms, $N_1=xN$ ($x\in [0,1]$) atoms in lattice $1$ with specific area $a_1$ and $N_2=(1-x)N$ atoms in lattice $2$ with specific area $a_2$, with the mean specific area $a_0=A/N$. The mean energy $E/N=xE^\text{coh}_1(a_1)+(1-x)E_2^\text{coh}(a_2)$ is minimized when
\begin{equation}
\frac{\partial E^\text{coh}_1(a_1)}{\partial a_1}=\frac{\partial E^\text{coh}_2(a_2)}{\partial a_2}\, ,
\label{eq:coexistence_condition}
\end{equation}
with $x=(a_0-a_2)/(a_1-a_2)$, provided that $\partial^2 E^\text{coh}_j(a_j)/\partial a_j^2>0$ for both $j=1,\,2$. Thus, lattices with specific areas $a_1$ and $a_2$ may coexist if their stresses are the same and cohesion energies are convex functions. As a result, the mean specific area $a_0=a_2+x(a_1-a_2)$ can have any value within $[a_1,a_2]$, including densities at which no single lattice alone would be stable (Fig.~\ref{fig:coexistence}a). This implies that, by lattice coexistence, a pore of \emph{area $A$ may stabilize metallene patches with considerably different numbers of atoms}. Based on cohesion energy curves, such coexistence is possible for most elements (Fig. S1) (ESI). We demonstrated the dynamical stability of coexisting hexagonal and buckled honeycomb lattices of Au in a $0.25$-ns molecular dynamics simulation with $2018$-atom supercell (Fig.~\ref{fig:coexistence}b) (ESI). The coexistence demonstration also involved a grain boundary, which behaved flexibly and did not interfere with the coexistence dynamics.
Stable lattice coexistence is not due to entropic effects, which anyway are more relevant in alloys and compounds, not elemental metallenes. This assertion aligns with experimental studies by Sharma \emph{et al.} \cite{Sharma}, where stable coexistence has been observed without resorting to entropic contributions.

Let us summarize our results to address the following question: \emph{which elements show the best overall potential for experimental stability?} The likelihood of experimental success increases upon increasing energetic and dynamical stability. Therefore, we rank the energetic stability of element $i$ using the quantity $
e_i=\sum_L^\text{lattices}\langle E^\text{coh}_L \rangle /(6 E^\text{coh}_{3D})$, where $\langle E^\text{coh}_L \rangle$ is the average cohesion energy of stable intervals of lattice $L$. This quantity will be close to zero with cohesion far from bulk and close to one if all lattices have stable intervals near bulk cohesion. Moreover, we rank the dynamical stability of element $i$ using the quantity $s_i=\sum_L^\text{lattices} \int_\text{stable} da'/\Delta a'$, where $\Delta a'=a'_\text{max}-a'_\text{min}=10.8$ is the maximum range for stable lattices (Fig.~\ref{fig:all_elements}). This quantity sums up all the stable intervals from all lattices and is zero for elements without any stable lattices and close to one for elements with stable lattices at all reasonable densities. Finally, we calculate the average rank for the element $i$ as $\mathcal{M}_i=(R_i^e+R_i^s)/2$, where $R_i^e$ is the rank of element $i$ in terms of energetic stability and $R_i^s$ is the rank of element $i$ in terms of dynamical stability. Note that our aim is to \emph{rank} elements, not to create a quantitative stability measure. 

\begin{figure}[b!]
\centering
\includegraphics[width=1.0\columnwidth]{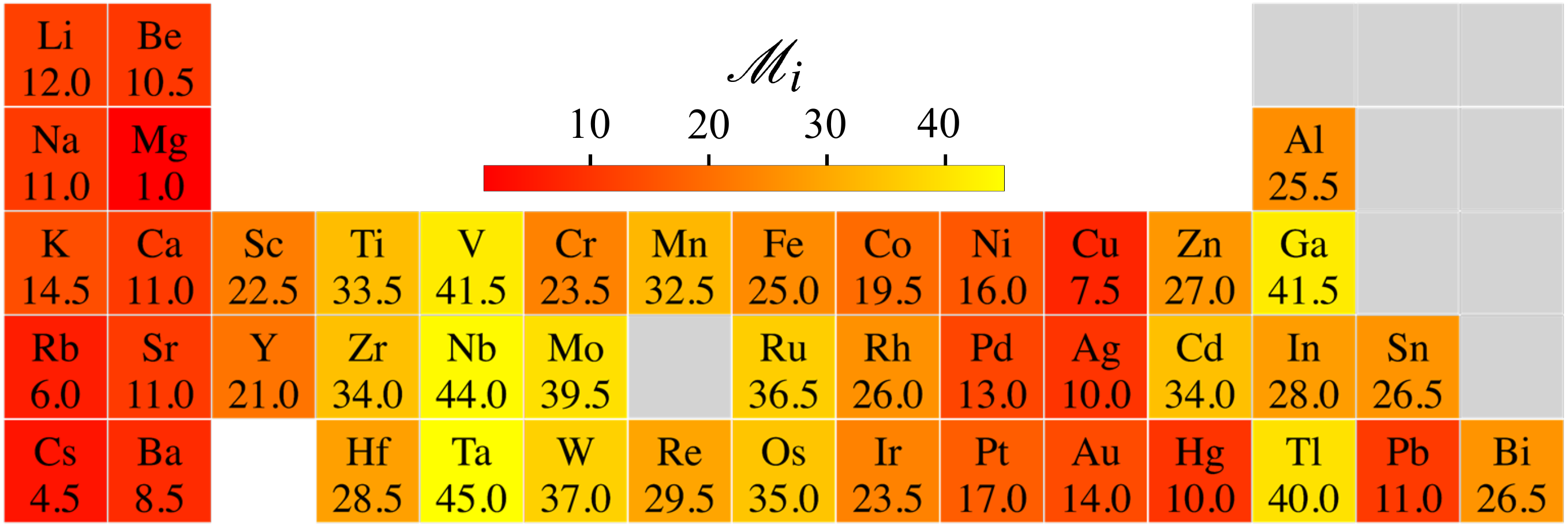}
\caption{Identifying elements with the best potential for 2D stability. Heatmap of the ranking parameter $\mathcal{M}_i$ characterizing the overall 2D stability of each element; lower $\mathcal{M}_i$ indicates better 2D stability.} 
\label{fig:FOM}
\end{figure}

The heatmap of $\mathcal{M}_i$ thus emphasizes the elements' potential for 2D stability from both energetic and dynamical viewpoints, compared to other elements (Fig.~\ref{fig:FOM}). The best potential is with alkali, earth alkali, and late-transition metals. Mg ranks first from both energetic and dynamic viewpoints. These trends agree with previous research \cite{review1, review2, review3, review4, elemental2D, review5}. Groups in the middle of the transition series rank the worst, Cr being a notable exception \cite{2DCr}. The trends also align with surface energies related to exfoliability and two-dimensional energetic stabilities \cite{surface, review5}. The experimentally observed Cr, Au, and Sn metallenes rank moderately among all metals \cite{2DAu, 2Dtin, 2DGold, 2DCr,zhao_free-standing_2014, review5}, which implies that several more 2D metals are awaiting experimental realization.

Our results should be approached with certain limitations in mind. They have limited applicability to substrate-supported monolayers, whose dynamical stabilities are likely dominated by the substrate \cite{10.1103/physrevlett.102.106102,Yin2015}. A higher level of theory might influence results quantitatively, but only to a minor degree, as suggested by earlier work \cite{OptimalDFT}. Finally, our approach is based on investigating the lattice properties with a \emph{local} viewpoint (Fig.~\ref{fig:systems}a), and it does not recognize finite-size effects, effects of interfaces to the covalent template, or potential lattice stabilization by superstructures such as charge density waves \cite{cdw_3}; these effects will be investigated in forthcoming studies.

To conclude, $45$ metallenes have $128$ dynamically stable 2D crystalline lattices at sporadic density intervals, the extent of which can be expanded by lattice coexistence. On average, metallenes show the best stability under gentle tensile strain. Therefore, given metallenes' ductility \cite{nevalaita_atlas_2018, plentyofmotion}, experiments should approach metallenes with a paradigm that abandons fixed lattice constants and considers them as yielding membranes that can be stabilized better under gentle tension. This paradigm may also help understand certain quantitative discrepancies between theory and the experimentally observed 2D patches \cite{zhao_free-standing_2014, 2DCr, 2Dtin}. The qualitative trends in the stability ranking of elements, which combined energetic and dynamic stability viewpoints, comply with the stability for experimentally observed metallenes (Fig.~\ref{fig:FOM}). At the same time, it recognizes several promising candidates, leaving plenty of room for new metallenes waiting for discovery.

\section*{Author contributions}
K. Raza Abidi: Investigation, Validation, Formal Analysis, Visualization, Methodology (equal), Funding Acquisition (equal), Writing – Original Draft; P. Koskinen: Conceptualization, Resources, Methodology (equal), Supervision, Funding Acquisition (equal), Project Administration, Writing – Review \& Editing

\section*{Conflicts of interest}
There are no conflicts to declare.

\section*{Acknowledgments}
We acknowledge the Vilho, Yrjö, and Kalle Väisälä Foundation of the Finnish Academy of Science and Letters and the Jane and Aatos Erkko Foundation for funding (project EcoMet) and the Finnish Grid and Cloud Infrastructure (FGCI) for computational resources.

\renewcommand\refname{References}

\end{document}